\documentclass[usenatbib]{mn2e}
\newcommand{\bib}{\bibitem[\protect\citeauthoryear}
\usepackage{epsf}
\begin{document}
\title[Synchrotron emission]{Synchrotron emission from 
anisotropic disordered magnetic fields}
\author[R.A. Laing]{R.A. Laing\thanks{E--mail: R.A.Laing@rl.ac.uk}\\
       Space Science and Technology Department, CLRC, Rutherford Appleton Laboratory,
       Chilton, Didcot, Oxfordshire OX11 0QX \\
       University of Oxford, Department of Astrophysics, Nuclear and
       Astrophysics Laboratory, Keble Road, Oxford OX1 3RH}

\date{Received }
\maketitle

\begin{abstract} 
We derive expressions for the total and linearly-polarized synchrotron
emissivity of an element of plasma containing relativistic particles and
disordered magnetic field which has been sheared or compressed along three
independent directions.  Our treatment follows that given by \citet{MS} in
the special case of a power-law electron energy spectrum.  We show that
the emissivity integrals depend on a single parameter, making it
straightforward to generate one-dimensional look-up tables.  We also
demonstrate that our formulae give identical results to those in the
literature in special cases.
\end{abstract}


\section{Introduction}
\label{Introduction} 

The inference of the three-dimensional structure of the magnetic field in
a source of synchrotron radiation from the observed emission at a single
frequency is not unique, but considerable progress can be made if
well-resolved images of total intensity and linear polarization are
available.  We have developed models of jets in low-luminosity radio
galaxies which predict the observed emission for comparison with
observations (\citealt{LPdRF}; Laing \& Bridle, in preparation). Our
fundamental assumption about the field structure is that it is random on
small scales, but made anisotropic by shear and compression.  Given the
large number of observed data points, the need to integrate along the line
of sight and the iteration required to optimize the fit between model and
data, we found that existing algorithms for the calculation of synchrotron
emissivity from disordered fields were too cumbersome.  In order to
minimize the resulting computation, we have developed an efficient method
of evaluating the Stokes parameters $I$, $Q$ and $U$ for synchrotron
radiation from electrons radiating in a three-dimensional, anisotropic
magnetic field.  This technique should be applicable to a variety of
astrophysical synchrotron sources and is the subject of the present paper.

The basic theory of synchrotron emission from an optically-thin power-law
energy distribution of relativistic electrons $n(E) dE \propto
E^{-(2\alpha+1)} dE$ is well-documented (e.g. \citealt{RL,BJA}).  The
spectrum of the emitted radiation also has a power-law form with spectral
index $\alpha$ ($S \propto \nu^{-\alpha}$). For a uniform field, $B$, at
an angle $\theta$ to the line of sight, the emissivity is $\propto
(B\sin\theta)^{1+\alpha}$ and the degree of polarization has its maximum
value $p_0 = (3\alpha+3)/(3\alpha+5)$.  The observed ${\bf E}$-vector is
perpendicular to the projected field direction on the sky in the absence
of propagation effects. \citet{S67} pointed out that the same results hold
for a one-dimensional disordered field with many reversals: $I$, $Q$ and
$U$ depend only on the orientation of the field, not on its vector
direction.

Expressions for synchrotron emission from a two-dimensional disordered
field with equal component variances along two orthogonal directions and
no component in the third were derived by \citet{L80}, and
discussed further by \citet{L81}.  The more general case of an
initially isotropic field compressed by an arbitrary factor along one
direction was considered by \citet{HAA}.  These
authors wrote down expressions for elements of field in terms of spherical
polar angles and averaged over the angles to derive the emitted Stokes
parameters. For a field structure described by one compression
factor, this approach is well suited to numerical calculation, but it
becomes extremely cumbersome if velocity shear is included.  
\citet[][hereafter MS]{MS} showed that a much simpler treatment
is possible if it is assumed that the initial field is isotropic with a
Gaussian distribution of field strengths.  Their formulae also include the
effects of inverse Compton and synchrotron losses, which cause a deviation
of the emitted spectrum from a power-law form. In this paper, we describe
a simplified version of their method, which applies to the
three-dimensional case, but is restricted to power-law spectra.  The
advantage is that computation of the emissivity is reduced to a
one-dimensional integration or table look-up.

The new formulae are derived in Section~\ref{3D}. In
Section~\ref{Compare}, we illustrate the use of the method in practice by
rederiving expressions in the literature for a number of special field
configurations.  Section~\ref{Summary} briefly summarizes our results.

\section{Computation of the Stokes parameters}
\label{3D}

\subsection{Assumptions}

The jet emission is taken to be optically-thin synchrotron radiation with
a power law frequency spectrum of index $\alpha$ as defined earlier.  We
make the conventional assumptions that the pitch-angle distribution of
radiating electrons is isotropic.  The observing frequency is taken to be
high enough that Faraday rotation may be neglected, or corrected
accurately.

The magnetic field is assumed to be disordered on small scales with zero
mean.  We and others have argued that this is likely to be a good
approximation for extragalactic radio sources, except perhaps on the
smallest linear scales (\citealt*{L81,BBR}; Laing \& Bridle, in
preparation).  We note, however, that if one of the three field components
is vector-ordered, then the results of our emissivity calculations are
unaltered.

\subsection{Geometry}

Following MS, the field structure is assumed to be generated from an
element of plasma containing radiating particles and an initially
isotropic field distribution by shear and compression.  We start with a
small cube of plasma whose sides are defined by three unit vectors in a
coordinate system $x$, $y$, $z$ moving with the emitting plasma (the
distinction between emitted and observed frames is important for
relativistic flow).  The coordinate axes are usually chosen
along obvious natural directions. This element of particles and field is then
compressed and sheared, forming a parallelepiped whose sides are defined
by the vectors ${\bf a}$, ${\bf b}$ and ${\bf c}$.  We then calculate the
components of these vectors in a coordinate system $X, Y, Z$ fixed in
space, with $Z$ along the line of sight.

MS show that the position angle $\chi_0$ of the apparent magnetic field
(perpendicular to the {\bf E}-vector for synchrotron radiation) is given
by:
\begin{equation}
\tan 2\chi_0 = \frac{2(a_X a_Y + b_X b_Y + c_X c_Y)}{a_X^2 - a_Y^2 + b_X^2
- b_Y^2 + c_X^2 - c_Y^2} \label{eq-chi}
\end{equation}
for any spectral index. $a_X$ is defined by:
\begin{equation}
a_X = {\bf a}.{\bf \bar{X}}/{\bf a}.{\bf b} \times {\bf c} 
\end{equation}
where ${\bf \bar{X}}$ is a unit vector along the X-axis, and so on. 
We will refer to $a_X, a_Y, b_X, b_Y, c_X, c_Y$ collectively as the edge
vector components.  We rotate about the line of sight by an
angle $\chi_0$ into a coordinate system $\hat{X}$, $\hat{Y}$, $\hat{Z}$
($\hat{Z} = Z$) in which Stokes $U = 0$. The rms field components along
the $\hat{X}$ and $\hat{Y}$ axes are $B_{\hat{X}}$ and $B_{\hat{Y}}$,
where:
\begin{eqnarray}
B_{\hat{X}}^2 & = & B_0^2 [ (a_X^2 + b_X^2 + c_X^2)\cos^2\chi_0 \nonumber \\
              & + & (a_Y^2 + b_Y^2 + c_Y^2) \sin^2\chi_0 \nonumber\\
      & + & 2(a_X a_Y + b_X b_Y + c_X c_Y) \sin\chi_0 \cos\chi_0]
              \label{eq-BX}\\  
B_{\hat{Y}}^2 & = & B_0^2 [ (a_X^2 + b_X^2 + c_X^2)\sin^2\chi_0 \nonumber\\
      & + & (a_Y^2 + b_Y^2 + c_Y^2) \cos^2\chi_0 \nonumber\\
      & - & 2(a_X a_Y + b_X b_Y + c_X c_Y) \sin\chi_0 \cos\chi_0]
              \label{eq-BY}\\ \nonumber  
\end{eqnarray}
Here, $B_0$ is the rms field along any direction for the initial isotropic
field.  Note that MS, in their equation (3.8), use the mean square value
of the total field ($3B_0^2$ in our notation).

\subsection{Gaussian field distributions}

For a power-law electron energy distribution, the observed emissivity from
a element of field of strength $B$ is always of the form $B^{1+\alpha}
\times$ a geometrical factor which is independent of $B$.  The simplest
initial field distribution is one where the field has constant intensity
and random direction. Whilst this gives simple expressions for the
emissivity if $\alpha = 1$, the more general case is extremely messy. MS
showed that a much simpler treatment is possible if the initial
distribution of field strengths is isotropic and Gaussian, with
probability distribution function (PDF):
\begin{eqnarray}
\lefteqn{F(B_x, B_y, B_z)dB_x dB_y dB_z} \nonumber \\ 
& = & (2\pi B_0^2)^{-3/2} \exp
\left (-\frac{B^2}{2B_0^2}\right ) dB_x dB_y dB_z \\\nonumber  
\end{eqnarray}
where $B_0$ is again the rms field in one coordinate, $B_x$, $B_y$ and $B_z$ are
the fields along $x$, $y$ and $z$ and $B^2 = B_x^2 + B_y^2 + B_z^2$.

In this case, the PDF of the field after shear or compression can still be
described by a multivariate Gaussian function.  The PDF of the components
of magnetic field perpendicular to the line of sight (responsible for
synchrotron emission) then has its major and minor axes along $\hat{X}$
and $\hat{Y}$, with the rms fields $B_{\hat{X}}$ and $B_{\hat{Y}}$ given
earlier.  Unless $B_{\hat{Y}} = 0$, the joint PDF for $B_\perp$ and the
position angle $\chi$ is as given by MS:
\begin{eqnarray}
\lefteqn{F_{\chi\perp} (B_\perp,\chi) dB_\perp d\chi} \nonumber \\  
& = & \frac{ B_\perp}{2\pi B_{\hat{X}}B_{\hat{Y}}}
\exp (-u B_\perp^2 - vB_\perp^2\cos 2\chi) dB_\perp d\chi \label{bivar}\\ \nonumber 
\end{eqnarray}
where 
\begin{eqnarray}
 u & = &  \frac{1}{4}
\left( \frac{1}{B_{\hat{X}}^2} + \frac{1}{B_{\hat{Y}}^2}\right) \\
 v & = & \frac{1}{4}
\left(\frac{1}{B_{\hat{X}}^2} - \frac{1}{B_{\hat{Y}}^2}\right) \\
\nonumber
\end{eqnarray}
The PDF for $B_\perp$ is given by integration of equation~(\ref{bivar})
over $\chi$:
\begin{eqnarray} 
F_\perp (B_\perp) & = & \frac{B_\perp
\exp(-u B_\perp^2)}{2\pi B_{\hat{X}}B_{\hat{Y}}} \int^{2\pi}_0 \exp(-v
B_\perp^2\cos 2\chi) d\chi \nonumber\\ 
& = & \frac{B_\perp
\exp(-u B_\perp^2)I_0(-vB_\perp^2)}{B_{\hat{X}}B_{\hat{Y}}}\\ \nonumber 
\end{eqnarray}
$I_n$ is the modified Bessel function of order n and we have used the
integral representation given by \citet{AS}, equation (9.6.16). This is
identical to equation (3.12) of MS, since $I_0$ is an even function.

\subsection{Synchrotron emissivities for a Gaussian field distribution}

We work with total and polarized intensity functions $j_I$ and $j_P$
defined such that the rest-frame emissivities are given by $\epsilon_ I =
k(n, \alpha) j_I$ and $\epsilon_P = k(n, \alpha) j_P$ where $k(n, \alpha)$
is proportional to the particle density $n$ and the functions are
normalized to have unit value for a uniform field of unit strength perpendicular to the line
of sight.  $j_I$ and $j_P$ are then functions of the field strength,
geometry and spectral index:
\begin{eqnarray}
j_I & = & \int^\infty_0 B_\perp^{1+\alpha}F_\perp (B_\perp) dB_\perp
   \nonumber\\ 
   & = & \frac{1}{B_{\hat{X}}B_{\hat{Y}}}
\int^\infty_0 B_\perp^{2+\alpha}\exp(-uB_\perp^2)I_0(-vB_\perp^2) dB_\perp \\
j_P & = & p_0 \int^\infty_0 dB_\perp \int^{2\pi}_0 d\chi 
B_\perp^{1+\alpha} \cos 2\chi F_{\chi\perp} (B_\perp,\chi) \nonumber \\ 
   & = & \frac{p_0}{B_{\hat{X}}B_{\hat{Y}}}  
\int^\infty_0 B_\perp^{2+\alpha}\exp(-uB_\perp^2)I_1(-vB_\perp^2) dB_\perp \\ \nonumber
\end{eqnarray}
We have again used the integral representation for $I_1$ given by
\citet{AS}, equation (9.6.19). 
It is convenient to write these formulae in terms of  the 
dimensionless variables $q = B_\perp/B_{\hat{X}}$  
and $\eta =B_{\hat{Y}}/B_{\hat{X}}$ (the axial ratio of the field PDF) and to 
define the functions $G_0(q) = \exp(-q) I_0(q)$ and 
$G_1(q) = \exp(-q) I_1(q)$.   For $q \gg n$, 
\[ I_n(q) \approx  \frac{1}{\sqrt{2\pi q}} \exp(q) \]  
so $G_n(q) \propto q^{-1/2}$.  The total and polarized intensity functions
then become:
\begin{eqnarray}
j_I & = & \frac{B_{\hat{X}}^{1+\alpha}}{\eta}\nonumber \\
    & \times & \int^{\infty}_{0}q^{2+\alpha}\exp(-q^2/2)
G_0\left[\frac{q^2}{4}\left(\frac{1}{\eta^2}-1\right)\right] dq \label{jI-gen}\\
j_P & = & \frac{p_0B_{\hat{X}}^{1+\alpha}}{\eta} \nonumber \\
 & \times & \int^{\infty}_{0}q^{2+\alpha}\exp(-q^2/2)
G_1\left[\frac{q^2}{4}\left(\frac{1}{\eta^2}-1\right)\right] dq \label{jP-gen}\\ \nonumber
\end{eqnarray}
In the special case $B_{\hat{Y}}= 0$ ($\eta = 0$), the PDF for $B_\perp$ is 
a one-dimensional Gaussian function:
\begin{equation}
 F_\perp(B_\perp) dB_\perp = \left (\frac{2}{\pi B^2_{\hat{X}}}\right)^{1/2} \exp \left (-\frac{B_\perp^2}{2B_{\hat{X}}^2} \right ) dB_\perp
\end{equation}
and the integral for $j_I$ can be
solved explicitly: 
\begin{eqnarray}
j_I &  = & B_{\hat{X}}^{1+\alpha} \pi^{-1/2} 2^{(1+\alpha)/2} 
\Gamma (1+\alpha/2) \label{jI-eta0}\\
j_P &  = & p_0 j_I \label{jP-eta0}\\ \nonumber
\end{eqnarray}
using the integral representation of the Gamma function, $\Gamma(w)$, given
in \citet[][hereafter
GR]{GR}\,8.310.  In this situation, there is no field component along the
$\hat{Y}$ axis and the degree of polarization takes its maximum value.

If $\alpha = 1$, there are simple analytical solutions.  The integrals
for $\eta \neq 0$ can be written in the form given by 
GR\,6.623 using the relations $I_0(w) =
J_0(iw)$ and $I_1(w) = -iJ_1(iw)$ and the substitution $w = q^2$.  The results 
are:
\begin{eqnarray}
j_I & = & B_{\hat{X}}^2 (1+\eta^2) \label{jI-alpha1}\\
j_P & = & p_0 B_{\hat{X}}^2 (1-\eta^2) \label{jP-alpha1}\\ \nonumber
\end{eqnarray}
and remain valid if $\eta = 0$. The degree of polarization is simply:
\begin{equation}
p = j_P/j_I = p_0 \left (\frac{1-\eta^2}{1+\eta^2}\right) \label{p1} 
\end{equation}

Finally, the Stokes parameters in the fixed $X$, $Y$ coordinate system are
given by:
\begin{eqnarray}
j_Q & = & j_P \cos 2\chi_0 \label{eqQ} \\
j_U & = & j_P \sin 2\chi_0 \label{eqU} \\ \nonumber
\end{eqnarray}
Note that these definitions of $Q$ and $U$ refer to the direction of the
apparent {\em magnetic} field, and that the position angle of the synchrotron
{\bf E}-vector is $\chi_0 + \pi/2$.

\subsection{Emissivities for an initial field of constant magnitude}
\label{Uni-field}

An initially isotropic Gaussian field with component rms $B_0$ can be
regarded as the superposition of a set of isotropic fields of constant
magnitude $B$ with a distribution:
\begin{equation}
F_{\rm 3D}(B) dB = \frac{2}{(2\pi)^{1/2}B_0^3}B^2\exp(-B^2/2B_0^2) dB 
\end{equation}
Consider two cubes, both
containing initially isotropic fields, but with different
distributions. The first has a Gaussian distribution, with component rms
$B_0$, as before. The second has a field of constant strength $B_0$. They
are stretched and sheared in identical fashions. Since the geometrical
factors in the emissivity calculation are independent of field strength,
the ratio of emissivities (Gaussian/constant) is:
\begin{eqnarray}
r & = & \frac{1}{B_0^{1+\alpha}}  \int^\infty_0 B^{1+\alpha} F_{\rm 3D}(B) dB
        \nonumber \\ 
        & = & 2^{(5+\alpha)/2}\pi^{-1/2}
\int^\infty_0\exp(-t^2)t^{3+\alpha} dt \nonumber \\
        & = & 2^{(3+\alpha)/2}\pi^{-1/2}
\Gamma (2 + \alpha/2) \label{eq-r}\\ \nonumber
\end{eqnarray} 
again using GR\,8.310.

The total and polarized emissivities for the case of a constant initial
field are $j_I/r$ and $j_P/r$, respectively.  $r = 3$ if $\alpha = 1$,
reflecting the facts that the emissivity is proportional to the square of
the field and that the mean square fields along a given direction are
$B_0^2$ and $B_0^2/3$ for the Gaussian and constant-field cases,
respectively.

\subsection{Implementation}

The integrals can be evaluated using standard routines, for example {\sc
qromo} and {\sc midexp} from \citet{NR}. These implement Romberg
integration for an integrand which decreases exponentially and an infinite
upper integration limit.  The functions $G_0$ and $G_1$ can be computed by
slightly modified versions of the subroutines {\sc bessi0} and {\sc
bessi1}, also from \citet{NR}. The emissivity functions depend only on the
ratio of the field rms's, $\eta$, for a given spectral index. Look-up
tables can be generated as functions of $\eta$ at the beginning of a
computation and thereafter polynomial interpolation can be used to derive
values as required.

\begin{figure}
\epsfxsize=8.5cm
\epsffile{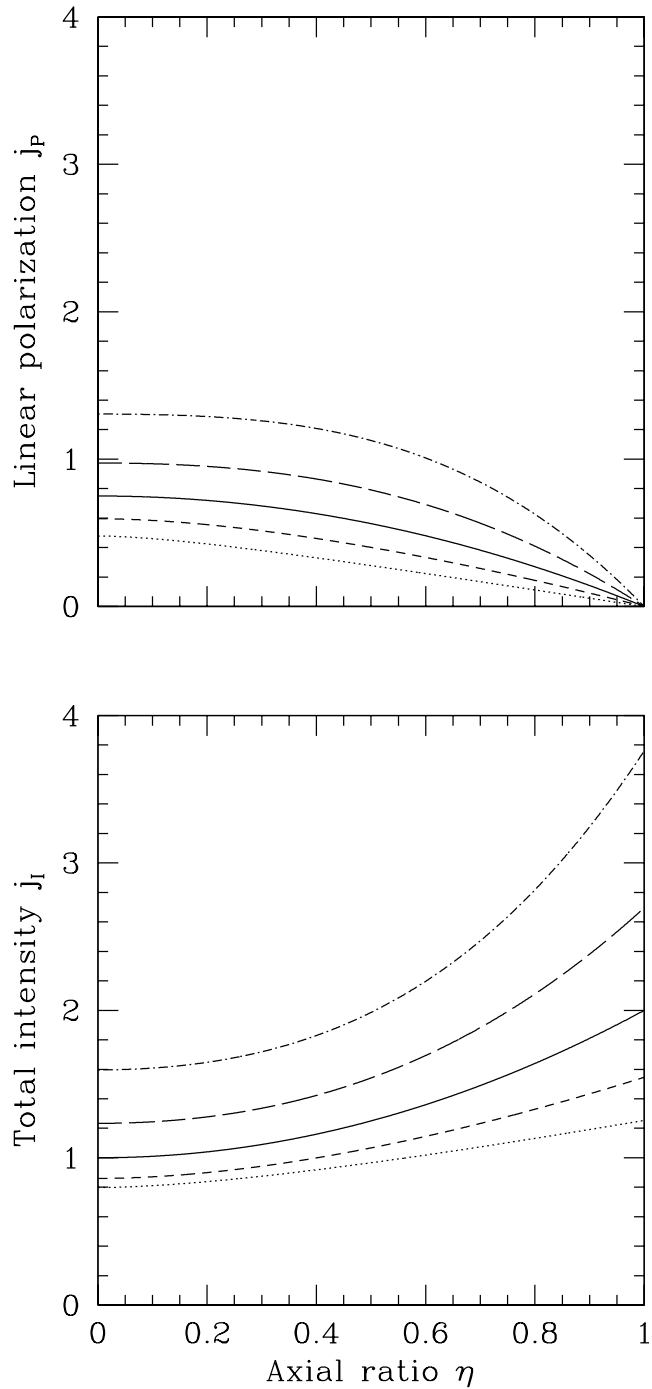}
\caption{Plots of the emissivity functions for total intensity (lower
panel) and linear polarization (upper panel) against the axial ratio of
the magnetic field PDF, $\eta$ for different spectral indices $\alpha$. 
The values of $\alpha$ are: 0.0 (dotted), 0.5 (short dashes), 1.0 (full
line), 1.5 (long dashes) and 2.0 (dash-dot).  \label{ipplot}}
\end{figure}

\begin{figure}
\epsfxsize=8.5cm
\epsffile{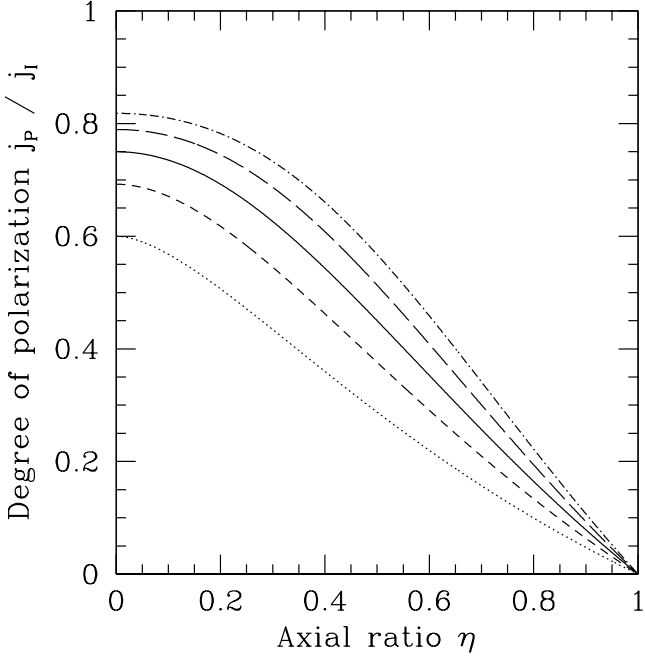}
\caption{Plot of the degree of polarization  against the axial ratio of
the magnetic field PDF, $\eta$ for different spectral indices $\alpha$. 
The values of $\alpha$ are: 0.0 (dotted), 0.5 (short dashes), 1.0 (full
line), 1.5 (long dashes) and 2.0 (dash-dot).  \label{degplot}}
\end{figure}

The emissivity functions $j_I$ and $j_P$ are plotted against axial ratio
$\eta$ in Fig.~\ref{ipplot} for a range of spectral indices typical of
optically-thin synchrotron radiation. The corresponding degrees of
polarization, $p = j_P/j_I$, are shown in Fig.~\ref{degplot}. Although the
variation of total intensity with spectral index is substantial, that of
the degree of polarization is not, and the analytical formula for $\alpha
= 1$ (equation~\ref{p1}) provides a good approximation for most sources,
as previously noted in less general cases by \citet{L80} and \citet{HAA}.

\section{Example calculations}
\label{Compare}

In this section, we outline the use of the method in practice and
demonstrate that the formulae given in the previous section are equivalent
to expressions in the literature for a number of simple field
distributions.  These distributions are non-Gaussian, and their emissivity
functions are denoted by $f_I$ and $f_P$, to distinguish them from the
Gaussian equivalents.

\subsection{General approach}
\label{gen}

In one class of problem, we are given the rms field components along three
orthogonal directions. Such a configuration can always be generated from a
unit cube containing an isotropic field by stretch or compression along
the three axes.  We can choose orthogonal vectors ${\bf a}$, ${\bf b}$ and
${\bf c}$ along the axes so that the volume $V$ of the resulting cuboid
has any specified value.  The initial field $B_0$ and the magnitudes of the
vectors are determined by:
\begin{eqnarray}
\langle B_x^2 \rangle^{1/2}& = &aB_0/V \label{comp1}\\
\langle B_y^2 \rangle^{1/2}& = &bB_0/V \label{comp2}\\
\langle B_z^2 \rangle^{1/2}& = & cB_0/V \label{comp3}\\
V & = & abc \label{vol}\\ \nonumber
\end{eqnarray}
so we can solve for $B_0$, $a$, $b$ and $c$.

We can extend this approach to follow the emission from an element of
fluid in a prescribed flow pattern (including the effects of shear) by
tracking the vectors ${\bf a}$, ${\bf b}$ and ${\bf c}$ as in MS, using
the values given above as initial conditions.

\subsection{One-dimensional fields}

To get a one-dimensional field, we start with a cube of initially
isotropic, Gaussian field of rms $B_0$ in each coordinate and expand it by
a factor $c \rightarrow \infty$ in one direction ($z$), leaving the other
two dimensions constant. The rms field in the $z$ direction remains $B_0$,
whilst those in the other two coordinates $\rightarrow 0$.  Suppose that
the $z$-axis makes an angle $\theta$ with the line of sight. In the
formulation of Section~\ref{3D}, $\eta = 0$ and the only non-zero edge
vector component is $c_X = \sin\theta$.  The emissivities are given by
equations~(\ref{jI-eta0}) and (\ref{jP-eta0}) with $B_{\hat{X}} =
B_0\sin\theta$.

Alternatively, we can use the standard emissivity functions for a
one-dimensional field of strength $B$:
\begin{eqnarray}
f_I & = & (B\sin\theta)^{1+\alpha} \\
f_P & = &p_0 f_I \\ \nonumber
\end{eqnarray} 
together with the PDF for  the total field, which is a
one-dimensional Gaussian function:
\begin{equation}
 F_{\rm 1D}(B) dB = \left (\frac{2}{\pi B_0^2}\right)^{1/2} \exp \left (-\frac{B^2}{2B_0^2} \right ) dB
\end{equation}
Integration of $B^{1+\alpha}$ over the Gaussian PDF gives
equations~(\ref{jI-eta0}) and (\ref{jP-eta0}) with  $B_{\hat{X}} =
B_0\sin\theta$.

\subsection{Two-dimensional fields}

Analytical expressions for the total and polarized emissivities for a
sheet of disordered field with no component in one direction but equal
components in the other two coordinates were given by \citet{L80}, who
considered a field of constant strength $B$ and random direction, confined
to a plane (note that this distribution is not produced by simply 
squashing a three-dimensional field of constant magnitude and random
direction).  

In order to compare these formulae with their equivalents from
Section~\ref{3D}, we consider a cube of  initially isotropic, Gaussian field of
rms $B_0$ in each coordinate which is expanded by factors $a$, $a$ and
$1/a$ along the $x$, $y$ and $z$ axes in the limit $a \rightarrow
\infty$, so that there is no field in the $z$ direction, but the rms
values in the $x$ and $y$ directions remain equal to $B_0$.  Since the
$z$-axis makes an angle $\theta$ with the line of sight, the edge vector
components are:
\[
a_X  =  1 ,\:
b_X   =   0 , \:
c_X   =   0
\]
\begin{equation}
a_Y   =   0 ,\:
b_Y   =   \cos\theta  ,\:
c_Y   =  0
\end{equation}
so $B_{\hat{X}} = B_0$ (since the field component along the major axis is
not projected) and $\eta = \cos\theta$.  The emissivity functions are
given by inserting these values into equations~(\ref{jI-gen}),
(\ref{jP-gen}), (\ref{jI-eta0}) and (\ref{jP-eta0}). For $\theta \neq
\pi/2$, we make the substitutions:
\begin{eqnarray} 
t & = & \frac{q^2}{4}  \left (\frac{1}{\eta^2}-1 \right ) \\
w & = & \frac {1+\cos^2\theta}{2\cos\theta} \\ \nonumber
\end{eqnarray}
into the expressions for $j_I$ and $j_P$, so 
\begin{eqnarray}
j_I & = & B_0^{1+\alpha}2^{2+\alpha}
\frac{\cos^{2+\alpha}\theta}{\sin^{3+\alpha}\theta} \nonumber \\
& \times & \int^\infty_0
t^\frac{1+\alpha}{2} \exp [-tw(w^2-1)^{-1/2}]I_0(t) dt \\
j_P & = & p_0 B_0^{1+\alpha}2^{2+\alpha}
\frac{\cos^{2+\alpha}\theta}{\sin^{3+\alpha}\theta} \nonumber \\
& \times &\int^\infty_0
t^\frac{1+\alpha}{2} \exp [-tw(w^2-1)^{-1/2}]I_1(t) dt \\ \nonumber
\end{eqnarray}
The integrals can be written in terms of Associated Legendre and Gamma
functions using GR\,6.624 and 8.731:
\begin{eqnarray}
j_I & = & B_0^{1+\alpha}2^\frac{1+\alpha}{2} \Gamma \left
(\frac{3+\alpha}{2} \right )(\cos\theta)^{\frac{\alpha+1}{2}}\nonumber \\
& \times & P_\frac{\alpha+1}{2}\left ( \frac{1+\cos^2
\theta}{2 \cos\theta} \right ) \label{2DI} \\
j_P & = & p_0 B_0^{1+\alpha} 2^\frac{1+\alpha}{2} \Gamma \left
(\frac{3+\alpha}{2} \right )(\cos\theta)^{\frac{\alpha+1}{2}} \left
(\frac{2}{1+\alpha} \right ) \nonumber \\
& \times & P^1_\frac{\alpha+1}{2}\left ( \frac{1+\cos^2
\theta}{2 \cos\theta} \right ) \label{2DP}\\ \nonumber
\end{eqnarray}
unless $\theta = \pi/2$, in which case equations~(\ref{jI-eta0}) and
(\ref{jP-eta0}) apply, with $B_{\hat{X}} = B_0$.

For a spectral index $\alpha = 1$,
equations~(\ref{jI-alpha1}) and (\ref{jP-alpha1}) give:
\begin{eqnarray}
j_I & = & B_0^2 (1 + \cos^2\theta) \label{2DI1} \\
j_P & = & p_0 B_0^2 \sin^2\theta \label{2DP1} \\ \nonumber
\end{eqnarray}

Alternatively, we can start from the expressions given by \citet{L80}
and average over the PDF for the total field. We make three modifications to the notation:
\begin{enumerate}
\item We normalize over the PDF for azimuthal angle, introducing an extra
factor of $2\pi$.
\item We use the angle $\theta$ between the normal to the field sheet and
the line 
of sight instead of $\beta = \pi/2 - \theta$.
\item We make use of GR\,8.731 to simplify the expression for polarized
emissivity.
\end{enumerate}
The emissivity functions become:
\begin{eqnarray}
f_I(\theta) & = & B^{1+\alpha}(\cos\theta)^{\frac{\alpha+1}{2}} P_\frac{\alpha+1}{2}\left ( \frac{1+\cos^2
\theta}{2 \cos\theta} \right )\\
f_P(\theta) & = & p_0 B^{1+\alpha} (\cos\theta)^{\frac{\alpha+1}{2}}
\nonumber \\
& \times&  \left
(\frac{2}{1+\alpha} \right ) P^1_\frac{\alpha+1}{2}\left ( \frac{1+\cos^2
\theta}{2 \cos\theta} \right )\\ \nonumber
\end{eqnarray}
For an edge-on field
sheet ($\theta = \pi/2$), the argument of the Legendre functions 
$\rightarrow \infty$ and the expressions are replaced by:
\begin{eqnarray}
f_I(\pi/2) & = & B^{1+\alpha} \pi^{-1/2} \Gamma (1
+\frac{\alpha}{2})/\Gamma (\frac{3+\alpha}{2}) \\
f_P & = & p_0 f_I \\ \nonumber
\end{eqnarray}
If $\alpha = 1$, 
\begin{eqnarray}
f_I & = & B^2 (1 + \cos^2\theta) / 2 \\
f_P & = & p_0 B^2 \sin^2\theta / 2 \\ \nonumber
\end{eqnarray}

The total
field this time has a two-dimensional Gaussian PDF:
\begin{equation}
 F_{\rm 2D}(B) dB = \frac{B}{B_0^2} \exp \left (- \frac{B^2}{2B_0^2}
 \right ) dB 
\end{equation}
We can calculate the emissivities for this Gaussian field distribution by
integrating $B^{1+\alpha}$ over the PDF:
\begin{equation}
 \int^\infty_0 B^{1+\alpha} F_{\rm 2D}(B) dB = B_0^{1+\alpha}
2^\frac{1+\alpha}{2} \Gamma \left (\frac{3+\alpha}{2} \right )
\end{equation}
(GR\,8.310), so:
\begin{equation}
j_I/f_I = j_P/f_P  = 2^\frac{1+\alpha}{2} \Gamma \left (\frac{3+\alpha}{2} \right )
\end{equation}
If $\alpha = 1$, this ratio is $2 \Gamma (2) = 2$. The expressions are
indeed identical to those in equations~(\ref{2DI}) -- (\ref{2DP1}).

\subsection{Fields compressed along one axis}

A slightly more complex field configuration was studied by \citet{HAA}
in the context of shocks propagating through a jet. We calculate its
emissivity in this section in order to illustrate the use of the formalism
developed earlier for a three-dimensional case.  \citet{HAA} considered a
cube containing an initially isotropic field which is compressed to a 
factor $k$ times its original length along one axis ($z$) and observed in a frame where the
plane of compression makes an angle $\epsilon$ with respect to the line of
sight. We take the $X$ axis to be in the plane of compression, in which
case the vector components are:
\[
a_X  =  1/k ,\:
b_X   =   0 , \:
c_X   =   0
\]
\begin{equation}
a_Y   =   0 ,\:
b_Y   =   \sin\epsilon/k  ,\:
c_Y   =   \cos\epsilon
\end{equation}
The apparent field position angle $\chi_0 = 0$, so $\hat{X} = X$ and
$\hat{Y} = Y$. The field components and axial ratio are:
\begin{eqnarray}
B_{\hat{X}}^2 & = & B_0^2/k^2 \\ 
B_{\hat{Y}}^2 & = & B_0^2 (\sin^2\epsilon/k^2 + \cos^2\epsilon) \\
\eta & = & (\sin^2\epsilon + k^2\cos^2\epsilon)^{1/2} \\ \nonumber
\end{eqnarray}

The emissivity functions for $\alpha = 1$ and the degree of polarization,
$p$, are given by equations~(\ref{jI-alpha1}) and (\ref{jP-alpha1}), with
$r = 3$ (equation~\ref{eq-r}):
\begin{eqnarray} 
f_I & = & \frac{B_0^2}{3k^2}[2 - (1-k^2)\cos^2\epsilon] \\
f_P & = & p_0 \frac{B_0^2}{3k^2}(1-k^2)\cos^2\epsilon \\
p   & = & \frac{ p_0 (1-k^2)\cos^2\epsilon}{2 - (1-k^2)\sin^2\epsilon} \\
\nonumber 
\end{eqnarray}
The last expression is as given by \citet{HAA}.  We have also verified
that our numerical results (equations~\ref{jI-gen} -- \ref{jP-eta0}) are
identical to those of \citet{HAA} for $\alpha \neq 1$.

\section{Summary}
\label{Summary}

We have derived expressions for the synchrotron emissivity in Stokes
parameters $I$, $Q$ and $U$ for a three-dimensional, disordered,
anisotropic field.  These are computationally straightforward, requiring
only two one-dimensional integrations, which can in turn be used to
generate look-up tables.  In the special case $\alpha = 1$, the
expressions reduce to simple analytical formulae.

The steps in a practical calculation are as follows:
\begin{enumerate}
\item Start with a cube containing isotropic field, either of constant
magnitude $B_0$, or with a Gaussian distribution of component rms $B_0$.
\item Compress and/or shear the cube to get the prescribed field configuration
(Section~\ref{gen}; equations~\ref{comp1} -- \ref{vol}). 
\item Evaluate the components of the 3 vectors defining the edges of the
resulting parallelepiped in the observed coordinate system and normalize
them by the volume to give the edge vector components $a_X, a_Y, b_X, b_Y,
c_X, c_Y$.
\item Work out the apparent field position angle $\chi_0$ from
equation~(\ref{eq-chi}).
\item Derive the field components along the principal axes, and their
ratio $\eta$ (equations~\ref{eq-BX} and \ref{eq-BY}).
\item Calculate the total and polarized emissivities.  The relevant pairs
of equations for a Gaussian initial field strength
distribution are summarized in Table~\ref{eq-table}.
\item For a constant initial field strength, divide by the factor $r$
given in equation~(\ref{eq-r}).
\item Rotate by $\chi_0$ to obtain observed Stokes $Q$ and $U$
(equations~\ref{eqQ} and \ref{eqU}).
\end{enumerate}

\begin{table}
\begin{center}
\caption{Equations for total and polarized emissivity functions for the
cases considered in the text.\label{eq-table}}
\begin{tabular}{ccc}
&&\\
\hline
&&\\
\multicolumn{2}{c}{$\alpha \neq 1$} & $\alpha = 1$ \\
 $\eta \neq 0$ & $\eta = 0$ & any $\eta$ \\
&&\\
\hline
&&\\
\ref{jI-gen}, \ref{jP-gen} & \ref{jI-eta0}, \ref{jP-eta0} & \ref{jI-alpha1}, \ref{jP-alpha1} \\
&&\\
\hline
\end{tabular}
\end{center}
\end{table}

We have shown that the formulae reduce to the standard expressions in the
literature for one- and two-dimensional fields and for an initially isotropic
field compressed along one axis.

\section*{Acknowledgments}

This paper has made extensive use of routines from Numerical Recipes 
\citep{NR}, which is gratefully acknowledged.

\end{document}